\documentstyle[12pt]{article}

\newcommand{\be}{\begin{equation}}
\newcommand{\ee}{\end{equation}}
\newcommand{\bea}{\begin{eqnarray}}
\newcommand{\eea}{\end{eqnarray}}
\newcommand{\ba}{\begin{array}}
\newcommand{\ea}{\end{array}}     
\newcommand{\nn}{\nonumber \\}
\newcommand{\half}{\frac{1}{2}}
\newcommand{\bb}{\bibitem}

\begin{document}
\begin{titlepage}

\hspace{9.5cm}{IFUSP-P/1229}

\hspace{9.5cm}{hep-th/9608049}

\vspace{.5cm}
\begin{center}
\LARGE

{\sc BLACK HOLES IN THE GAUGE THEORETIC FORMULATION 
OF DILATONIC GRAVITY}

\vspace{.5cm}
\large

Marcelo M. Leite \footnote{e-mail: mml@if.usp.br} \\ and \\ Victor
O. Rivelles \footnote{e-mail: vrivelles@if.usp.br} 

\vspace{.5cm}

{\it Instituto de F\'\i sica, Universidade de S\~ao 
Paulo} \\ {\it C.Postal 66318, 05389-970, S.Paulo, SP, Brazil}

\vspace{.5cm}
August, 1996
\end{center}
\vspace{1cm}

\begin{abstract}
We show that two-dimensional topological BF theories coupled to
particles carrying non-Abelian charge admit a new coupling involving 
the Lagrange multiplier field. When applied to the gauge theoretic
formulation of dilatonic gravity it gives rise to a source term for
the gravitational field. We show that the system admits black hole
solutions. 
\end{abstract}
\end{titlepage}
\newpage

Two-dimensional gravity theories have been studied in the hope to have
a better understanding of the quantum properties of its
four-dimensional counterpart. Usually others fields are introduced if
we require non trivial dynamics for the gravitational field.  A 
class of interesting models introduces just one scalar field known as
the dilaton. 
A particular model where black holes can
be formed and the quantization can be performed is the 
Callan, Giddings, Harvey and Strominger (CGHS) model \cite{cghs}. It is
described by the action
\be
\label{cghs}
S = \int d^2x \, \sqrt{-\bar{g}} [ e^{-2\phi}  ( \bar{R} - 4
\bar{g}^{\mu\nu} \partial_\mu \phi \partial_\nu \phi - \Lambda )
+ \half \sum_i \bar{g}^{\mu\nu} \partial_\mu f_i \partial_\nu f_i
], 
\ee
where $\bar{R}$ is the curvature scalar build with the metric
$\bar{g}_{\mu\nu}$, $\phi$ is the dilaton, $\Lambda$ 
is the cosmological constant and $f_i$ is a set of scalar matter 
fields. 
If the conformal transformation 
$g_{\mu\nu} = e^{-2\phi} \bar{g}_{\mu\nu}$ 
is performed the action (\ref{cghs}) takes the form  \cite{verlinde}
\be
\label{string}
S = \int d^2x \, \sqrt{-g}\, ( \eta R - \Lambda  + \half \sum_i
{g}^{\mu\nu} \partial_\mu f_i \partial_\nu f_i ), 
\ee
where $\eta = e^{-2\phi}$ and $R$ is the scalar curvature built with the
metric $g_{\mu\nu}$. Now the field equation for $\eta$ implies that $R
= 0$ so that the two-dimensional space-time is locally flat and there
is no black hole solution. This happens because the conformal
transformation changes the geometry but since is just
a field redefinition the physical content of 
theory should not be affected by it. 
Usually the metric $\bar{g}_{\mu\nu}$,
which can describe black holes, is called the ``physical'' metric
while $g_{\mu\nu}$, which is locally flat, is called the ``stringy''
metric and physical interpretations depending on the space-time
geometry are usually taken using the ``physical'' metric. 
At the quantum level the conformal transformation is
more problematic. While the model described by (\ref{cghs}) presents
Hawking radiation the model described by (\ref{string}) has no Hawking
radiation. It has been argued that with proper care of the conformal
transformation no ambiguity exists  \cite{cadoni}.
Even so the quantization in either form is not free of troubles
\cite{jackiw1}.

An important aspect of the action (\ref{string}) is that it can be
rewritten as a topological gauge theory of the BF type  \cite{birm}
with a gauge group which is a central extension of the two-dimensional
Poincar\'e group  \cite{cangemi}. If we couple matter in this
formulation it should be coupled in a gauge invariant way. A proposal
to do that is to use a Higgs-like mechanism which introduces a new
field, the Poincar\'e coordinate, to describe point-like
matter \cite{grignami}. Another possibility makes use of a formulation of
relativistic particles which carry non-Abelian charges \cite{bal}. It
was applied to the gauge theoretic version of dilatonic gravity and some
solutions were shown to be equivalent to those of the Poincar\'e coordinate
formulation \cite{stephany}. An important feature of both approaches to
include matter is that the curvature equation $R=0$ never acquires a
source term. We will show that in the formulation where particles
carry non-Abelian charge a new gauge invariant coupling does exist for
topological BF theories. This new
coupling provides a source term for the curvature equation and black
hole solutions can then be found. 

Non-Abelian degrees of freedom for point particles were originally
introduced in the context of QCD \cite{wong,bal}. They are described by
the group element $g(\tau)$ and a real constant element of the algebra
$K$, $\tau$ being the proper time of the particle. It is useful to
introduce the variable 
\be
\label{Q}
Q(\tau) = g(\tau) K g^{-1}(\tau),
\ee
which is in the adjoint representation. The minimal coupling between
the particle and the gauge field can then 
be performed by introducing a covariant derivative
\be
\label{cov_derivative}
D_\tau = \frac{d}{d\tau} + e \dot{x}^\mu A_\mu( x(\tau) ). 
\ee
If we also consider a kinetic term for the relativistic particle then
an action which is gauge and reparametrization invariant is \cite{bal}
\be
\label{bal}
S = -m \int d\tau \, \sqrt{\dot{x}^2} + \int d\tau \, Tr( K
g^{-1}(\tau) D_\tau g(\tau) ).
\ee
This action is also invariant under the
transformation $K \rightarrow S K S^{-1}$ where $S$ is $\tau$
independent. This shows that the action (\ref{bal}) is independent of
the direction in the internal symmetry space 
given by $K$. 
Varying the action (\ref{bal}) with respect to $x^\mu(\tau)$ we get a
non-Abelian version of the Lorentz force
\be
\label{lorentz}
m \ddot{x}^\mu + \Gamma^\mu_{\nu\rho} \dot{x}^\nu \dot{x}^\rho = -e Tr(
F^\mu_{\,\,\,\,\nu} Q) \dot{x}^\nu,
\ee
while varying with respect to $g(\tau)$ we get a covariant conservation
equation for the non-Abelian charge $Q$
\be
\label{conservation}
\frac{dQ}{d\tau} + [ A_\mu(x(\tau)),
Q(\tau) ]   \dot{x}^\mu (\tau) = 0.
\ee
These equations are known as the Wong equations \cite{wong}. 

Consider now a two-dimensional BF topological field theory
\be
\label{bf}
S = \int d^2x \, Tr (\eta \, F),  
\ee
where $F = dA + A^2$ is the curvature two-form corresponding to the
connection one-form $A$ and $\eta$ is a zero-form transforming in the
co-adjoint representation of the gauge group. The action of
particles carrying non-Abelian charge (\ref{bal}) combined with the BF
action (\ref{bf}) was studied in  \cite{stephany} while the Abelian
case was treated in  \cite{lupi}. When compared with the pure BF case,
the main consequence of adding 
non-Abelian matter is that the field equation for the Lagrange
multiplier $\eta$ acquires a source term $\epsilon^{\mu\nu}D_\nu \eta
= J^\mu$
(where $J$ is the non-Abelian current of the particle) while the
connection remains flat $F = 0$. 

Since the structure of the BF theory requires two fields then, besides
the coupling involving the gauge field $A$, we can
consider another coupling involving the Lagrange multiplier 
$\eta$. A coupling of the type $Tr( \eta \, Q)$  is gauge invariant
but not  
proper time reparametrization invariant. In order to get a
reparametrization invariant action we introduce the worldline einbein
$e(\tau)$ and the respective mass term for the particle. So we can
consider an extension of the former actions to
\bea
\label{new}
S &=& \int d^2x \, Tr( \eta \, F) + \int d\tau \, Tr(g^{-1} K
\dot{g} ) + \nn 
 &+&  e \int d^2 x \int d\tau \, Tr( Q(\tau) A_\mu (x)) \,
\delta^2(x - x(\tau)) \, \dot{x}^\mu (\tau) + \nn
&+& g \int d^2x \int d\tau \, e(\tau) \, 
Tr( Q(\tau) \eta) \, \delta^2( x - x(\tau)) + \half m^2 \int d\tau \,
e(\tau), 
\eea
where $e$ and $g$ are independent coupling constants\footnote{In the
context of BF theories a kinetic term 
for the particle $ \int d\tau \,  \frac{1}{2e} \,g_{\mu\nu} \dot{x}^\mu
\dot{x}^\nu$ can also be added. However, in the gauge theoretic
formulation of dilatonic gravity the metric is composite of the gauge
fields 
and such a term would break the gauge invariance. For this reason we
do not include it in (\ref{new}).}. The action (\ref{new}) is invariant
under gauge transformations
\bea
\label{gauge-transf}
A &\rightarrow& h A h^{-1} - dh h^{-1}, \nn
\eta &\rightarrow& h \eta h^{-1}, \nn
g &\rightarrow& h g, \nn
K &\rightarrow& K,
\eea
and proper time reparametrization
\bea
\label{proper-time}
e^\prime(\tau^\prime) &=& \frac{d\tau}{d\tau^\prime} e(\tau), \nn
\dot{x}^{\prime\mu}(\tau^\prime) &=& \frac{d\tau}{d\tau^\prime}
\dot{x}^\mu(\tau), 
\eea
with all remaining fields being reparametrization scalars. 

The main consequence of the new coupling when compared to the former
case is that the connection is no longer flat and has as source the 
non-Abelian charge $Q$
\be
\label{f=q}
\epsilon^{\mu\nu} F_{\mu\nu} = g \int d\tau \, e(\tau) \,Q(\tau) \,
\delta^2(x - x(\tau)). 
\ee
As we shall see in the context of dilatonic gravity
theory this change allows us to find  black hole
solutions without making reference to any conformal transformation to
a ``physical'' metric. The field equations for the Lagrange multiplier
is 
\be
\label{deta}
\epsilon^{\mu\nu} D_\nu \eta = e \int d\tau \, Q(\tau) \, \delta^2
(x - x(\tau)) \, \dot{x}^\mu(\tau),
\ee
while the equation for non-Abelian charge is modified to 
\be
\label{dq}
\frac{dQ}{d\tau} + e [ A_\mu (x(\tau)), Q ] \dot{x}^\mu + g 
[ \eta (x(\tau)), Q] e(\tau) = 0.
\ee
generalizing the conservation equation (\ref{conservation}) to the BF
theory. The field equation obtained by varying the worldline einbein
is
\be
\label{einbein}
g \int d^2x \,\, Tr(Q(\tau) \eta(x) ) \, \delta^2(x - x(\tau) ) +
\half m^2 = 0.
\ee
Notice that the field equations obtained by varying $x^\mu (\tau)$
vanish identically. The reason is that by varying the action with
respect to $x^\mu(\tau)$ we get a contribution which is proportional to
the previous field equations (\ref{f=q}-\ref{einbein}) so that it
vanishes on-shell. This shows the topological character of the 
the non-Abelian particle in the sense that its local motion (described by
$x^\mu(\tau)$) is completely
arbitrary not being determined by any field equation. There are only
global restrictions to the motion of the particle as was shown
in  \cite{stephany,lupi} in the case $g = 0$. 

To consider the two dimensional dilatonic gravity theory we now choose
the gauge group as the central extension of the Poincar\'e
group \cite{cangemi}
\bea
\label{group}
{}[P_a, P_b] &=& \epsilon_{ab} Z, \nn
{}[J, P_a] &=& {\epsilon_a}^b P_b, \nn
{}[P_a, Z] &=& [J, Z] = 0, 
\eea
where $P_a$ is the translation generator, $J$ is the Lorentz transformation 
generator and $Z$ is a central element of the group. The
supersymmetric extension of 
(\ref{group}) was performed in \cite{me}. The flat Minkowski metric is
$h_{ab} =  
diag(-1, +1)$ and $\epsilon^{01} = 1$. When we consider the algebra
(\ref{group}) we can expand the one form gauge potential  
in terms of the generators of the algebra
\be
\label{potential}
A = e^a P_a + w J  + A Z.
\ee
The fields $e^a, w$ and $A$ are going to be identified with the 
zweibein, the spin connection and an Abelian gauge field, respectively. 
The Lagrange multiplier $\eta$ can be expanded as
\be
\label{eta}
\eta = \eta^a P_a + \eta_3 J + \eta_2 Z,
\ee
with components $\eta^a, \eta_2$ and $\eta_3$ with $\eta_2$ being
proportional to the dilaton in  (\ref{string}). 
Then the curvature two-form $F$ has components
\bea
\label{f}
F^a(P) &=& de^{a} + we^{b}\epsilon_{b}^{\,\,a}, \nn
F(J) &=& d\omega, \nn
F(Z) &=& dA + \half e^a e^b \epsilon_{ab}.
\eea
Similarly the non-Abelian charge $Q$ can be expanded as 
\be
\label{q}
Q = Q^a P_a + Q_3 J + Q_2 Z.
\ee

Then the field equations for the gauge fields (\ref{f=q}) are 
\bea
\label{f1}
& &\epsilon^{\mu\nu} ( \partial_\mu e_\nu^a + \omega_\mu e_\nu^b
\epsilon_b^{\,\,a} ) + g \int d\tau \, e(\tau) \, Q^a(\tau) \,
\delta^2(x - x(\tau)) = 0, \\
\label{f2}
& &\epsilon^{\mu\nu} \partial_\mu \omega_\nu + g \int d\tau \,
e(\tau) \, Q_3(\tau) \, \delta^2(x - x(\tau))  = 0, \\
\label{f3}
& &\epsilon^{\mu\nu} (\partial_\mu a_\nu + \half e_\mu^a e_\nu^b
\epsilon_{ab}) + g \int d\tau \, e(\tau) \, Q_2(\tau) \, \delta^2(x -
x(\tau)) = 0, 
\eea
while the field equations for the Lagrange multipliers (\ref{deta}) are
\bea
\label{eta2}
& &\epsilon^{\mu\nu} ( \partial_\nu \eta_a - \omega_\nu
\epsilon_a^{\,\,b} \eta_b + \eta_3 \epsilon_{ab} e_\nu^b ) 
+  e \int d\tau Q_a(\tau) \delta^2(x - x(\tau)) \dot{x}^\mu 
(\tau) = 0, \\
\label{eta1}
& &\epsilon^{\mu\nu} ( \partial_\nu \eta_2 + e_\nu^a \epsilon_a^{\,\,b}
\eta_b ) + e \int d\tau \,\, Q_2(\tau) \,\, \delta^2(x - x(\tau))
\dot{x}^\mu (\tau) =  0, \\ 
\label{eta3}
& &\epsilon^{\mu\nu} \partial_\nu \eta_3 +  e \int d\tau \,\,
Q_3(\tau) \,\, \delta^2(x - x(\tau)) \dot{x}^\mu (\tau) =  0.
\eea
The equations of motion for the non-Abelian charge (\ref{dq}) are
\bea
\label{q1}
& &\frac{dQ^a}{d\tau} + e \epsilon_b^{\,\,a} (e_\mu^b  
Q_3 - \omega_\mu Q^b ) \dot{x}^\mu + g \epsilon_b^{\,\,a} (\eta^b Q_3
- \eta_3 Q^b ) e(\tau) = 0, \\
\label{q3}
& &\frac{dQ_2}{d\tau} - \epsilon_{ab} (e e_\mu^a
Q^b \dot{x}^\mu - g \eta^a Q^b e(\tau) ) = 0, \\
\label{q2}
& &\frac{dQ_3}{d\tau} = 0.
\eea
Finally the equation of motion (\ref{einbein}) gives a constraint
among the 
non-Abelian charge, the Lagrange multiplier on the worldline and the
particle mass
\be
\label{einbein2}
g \int d^2x \, ( Q^a \eta_a + Q_3 \eta_2 + Q_2 \eta_3 ) \delta^2(x -
x(\tau))+ \half m^2 = 0. 
\ee
Due to the relation $\epsilon^{\mu\nu} \partial_\mu \omega_\nu =
\sqrt{-g} R$ it is now clear from (\ref{f2}) that in order to have a
non vanishing curvature $R$ we need to have $g \not= 0$. Notice also
that the two dimensional torsion $T_{\mu\nu}^a =  \partial_{[\mu}
e_{\nu]}^a + \omega_{[\mu} e_{\nu]}^b \epsilon_b^{\,\,a}$ may also be
present due to (\ref{f1}). 

We will now look for some solutions of the above equations. We will
find the general solution in the absence of matter and in the presence
of a non-Abelian point particle at rest. A more general analysis will
be done elsewhere. 
In order to solve (\ref{f1}-\ref{einbein2}) we have to perform
several gauge fixings and to choose a space-time trajectory for the
particle since, as remarked before, there is no field equation for
$x^\mu(\tau)$. We will look for static solutions so we use Rindler like 
coordinates $(x,t)$. For simplicity let us consider the proper time
gauge for the particle $e(\tau) = 1$ and set the particle at
rest in the origin
\be
\label{particle}
x(\tau) = 0, \qquad t(\tau) = \tau.
\ee
Let us consider first the gauge field sector. In the
gravitational sector we will choose a diagonal zweibein $e_0^1 = e_1^0
= 0$ with the non vanishing components satisfying $e_0^0 = ( e_1^1
)^{-1}$, 
and a vanishing space component of the connection $\omega_1 = 0$. For
the Abelian gauge field we will choose the axial gauge $A_1 = 0$. Then
eqs.(\ref{f1}-\ref{f3}) reduce to
\bea
\label{f_est1}
\partial_1 e_0^0 + \omega_0 e_1^1 &=& g Q^0 \, \delta(x), \\
\label{f_est2}
0 &=& g Q^1 \delta(x), \\
\label{f_est3}
\partial_1 \omega_0 &=& g Q_3 \, \delta(x), \\
\label{f_est4}
\partial_1 A_0 + e_0^0 e_1^1 &=& g Q_2 \, \delta(x).
\eea

If no matter is present ($ e = g = 0 $) then we find flat
space-time as the only solution. Explicitly we find 
\bea
\label{f_flat}
\omega_0 &=& -a, \quad e_0^0 = (b + 2ax)^\half, \nn
A_0 &=& -x + A, 
\eea
where $a, b$ and $A$ are integration constants. Notice that the line
element $ds^2 = -(b + 2ax) dt^2 + (b + 2ax)^{-1} dx^2$ has a singularity
at $x = - b/2a$. This coordinate singularity can be removed by the
coordinate transformation  
\bea
\label{coord_transf}
\sigma &=& \frac{1}{a} \sqrt{b + 2ax } \,\,\, \cosh(at), \nn
\tau &=&  \frac{1}{a} \sqrt{b + 2ax } \,\,\, \sinh(at)
\eea
for the patch $x>-b/2a, \sigma>0$ and similar transformations for the
three remaining patches. This coordinate transformation brings the
metric in (\ref{f_flat}) to its Minkowski form. The curvature scalar
vanishes dues to (\ref{f_est3}).  

Now consider the situation when matter is present. If $Q_3 = 0$ and
$Q^a \not= 0$ then the space-time has torsion but no curvature since 
(\ref{f_est1}) is proportional to the torsion. If
$Q_3 \not= 0$ and $Q^a = 0$ then the space-time has curvature but
no torsion. Let us consider the last case.  Take $Q_2$ and $Q_3$ as 
constants (as we shall see below $Q_2$ and $Q_3$ constants and $Q^a =
0$ is a solution of (\ref{q1}-\ref{q2}) ). Then we find as solution
of (\ref{f_est1}-\ref{f_est4}) 
\bea 
\label{schw}
\omega_0 &=& g Q_3 \epsilon(x), \quad e_0^0 = ( \tilde{b} - 2 g 
Q_3 |x| )^\half, \nn
A_0 &=& - x + g Q_2 \epsilon(x) + \tilde{A},
\eea
where $\tilde{b}$ and $\tilde{A}$ are integration constants. The
space-time described 
by (\ref{schw}) has a black hole \cite{mann} and the curvature scalar
is given by (\ref{f_est3}) $R = g Q_3 \delta(x)$. Notice that $g Q_3$
can now be understood as the black hole mass and it is essential to
have $g \not= 0$. 

In the Lagrange multiplier sector the gauge choices reduce
eqs.(\ref{eta2}-\ref{eta3}) to
\bea
\label{eta_est}
& & \partial_1 \eta_0 - \eta_3 e_1^1 = -e Q_0 \delta(x), \nn
& & \partial_1 \eta_1 = -e Q_1 \delta(x), \nn
& & \omega_0 \eta_1 = \omega_0 \eta_0 + \eta_3 e_0^0 = 0, \nn
& & \partial_1 \eta_2 - e_1^1 \eta_0 = -e Q_2 \delta(x), \nn
& & e_0^0 \eta_1 = 0, \nn
& &\partial_1 \eta_3 = -e Q_3 \delta(x).
\eea
In the absence of matter, using (\ref{f_flat}), we find the solution
\bea
\label{eta_flat1}
\eta_0 &=& \frac{\Lambda}{a} ( b + 2ax )^\half, \quad \eta_1 = 0, \\
\label{eta_flat2}
\eta_2 &=& \frac{\Lambda}{a} x + c, \quad \eta_3 = \Lambda, 
\eea
where $\Lambda$ and $c$ are integration constants. Notice that we have
identified one of the integration constants, the one coming from
integrating $\eta_3$ in (\ref{eta_est}), as the cosmological constant
$\Lambda$. The reason is that this solution in the absence of matter
\cite{cangemi} reproduces the results obtained in the original CGHS
formulation after performing the conformal transformation. The dilaton
(\ref{eta_flat2}) has an unusual form in the coordinates $(x,t)$. By
performing the coordinate transformation (\ref{coord_transf}) it
becomes proportional to $\sigma\tau$ assuming its usual form
\cite{cghs}. 

In the presence of matter with $Q_2$ and $Q_3$ constants and $Q^a = 0$
we find, using (\ref{schw}) 
\bea
\label{eta_schw}
\eta_0 &=& \frac{e}{g} (\tilde{b} - 2 g Q_3 |x|)^\half, \quad \eta_1 =
0, \nn 
\eta_2 &=& \frac{e}{g} x - e Q_2 \epsilon(x) + \tilde{c}, \quad
\eta_3 = - e Q_3 \epsilon(x), 
\eea
where $\tilde{c}$ is another integration constant and $\epsilon(x)$ is
the step function. The appearance of the step function in the solution
for the Lagrange multiplier fields signals that there are topological
restrictions to the motion of particles \cite{stephany,lupi}. It is
remarkable that the would be cosmological constant $\eta_3$ is now a
step function changing sign at the position of the particle. No
constant term is allowed in the solution for $\eta_3$ so if we set
$e=0$ then the cosmological constant vanishes. The
dilaton $\eta_2$ still has its linear term but has also acquired a
step function. We can however set $Q_2=0$ and still have a linear
dilaton and the black hole (\ref{schw}), which is independent of
$Q_2$. 

The equations for the non-Abelian charge
(\ref{q1}-\ref{q2}), after the gauge choice, reduce to
\bea
\label{q_est}
& &\dot{Q}^0 - e Q^1 \omega_0 - g ( Q^1 \eta_3 - Q_3 \eta^1
) = 0, \nn
& &\dot{Q}^1 - e ( Q^0 \omega_0 - Q_2 e_0^0 ) - g ( Q^0
\eta_3 - Q_3 \eta^0 ) = 0, \nn
& &\dot{Q}_2 - e Q^1 e_0^0 + g ( Q^0 \eta^1 - Q^1 \eta^0 ) = 0, \nn
& &\dot{Q}_3 = 0.
\eea
In the absence of matter these equations are trivially satisfied. In
the presence of matter with $Q^a = 0$ and $Q_2$ and $Q_3$ constantes
we find using (\ref{schw}) and  
(\ref{eta_schw}) that $Q_2$ and $Q_3$ are constants as we had
anticipated. 

Finally the constraint equation (\ref{einbein2}) becomes, after the
gauge choices, 
\be
\label{einbein3}
g [ Q^a(\tau) \eta_a(\tau,0) + Q_3(\tau) \eta_2(\tau,0) + Q_2(\tau)
\eta_3(\tau,0) ] + \half m^2 = 0.
\ee
It is trivially satisfied in the absence of matter. In the presence
of matter with $Q^a = 0$ (\ref{einbein3}) is ill defined since
according to (\ref{eta_schw}) $\eta_2$ and $\eta_3$ have a
discontinuity at $x=0$. We then take $Q_2 = 0$ and (\ref{einbein3})
becomes 
\be
\label{const_const}
g Q_3 \tilde{c} + \half m^2 = 0
\ee
giving a constraint among the integration constants $Q_3$ and
$\tilde{c}$ and the mass $m$. Recalling that the curvature scalar is
$R = g Q_3 \delta(x)$ we can interpret the black hole mass as being
due to the non-Abelian charge of the particle $Q_3$ or to its mass $m$
and to value of the dilaton at the point where the particle is
positioned. 

We have presented some local solutions for the gauge theoretic version
of dilatonic gravity theories with non-Abelian sources. We are still
investigating more general 
solutions and classifying them. Also it is necessary to perform a global
analysis of the solutions. When $g=0$ this can be done essentially
because the gauge connection is flat \cite{lupi}. In the present case
this task becomes more difficult since the gauge connection is no
longer flat due to (\ref{f=q}). Progress on these lines will be
reported elsewhere. 

VOR would like to acknowledge A. Restuccia for conversations and
J. Sthepany for discussions on an earlier version of this
work. VOR was partially supported by CNPq. MML was supported by
a grant from FAPESP.

\end{document}